\documentclass[
  ,draft            
  ]
  {aipproc}

\layoutstyle{6x9}

\begin{document}

\title{The brane as a Higgs domain wall: ideas and issues}


\author{Raymond R. Volkas}{
  address={School of Physics, Research Centre for High Energy Physics, 
The University of Melbourne, Victoria 3010 Australia}
}

\begin{abstract}
The most obvious field-theoretic model for a brane is a scalar
field domain wall or kink.  I discuss how this idea can be connected
with spontaneous internal symmetry breaking via a mechanism called
the ``clash of symmetries''.  Compatibility with Randall and Sundrum's
warped metric alternative to compactification is then demonstrated.
I end with brief remarks about open questions.
\end{abstract}

\maketitle


\section{Background}

In field theory, a brane can be inserted into a higher-dimensional action
as a fundamental object, or it can emerge dynamically as a scalar-field 
soliton \cite{rub}.  In this talk, I shall discuss the latter hypothesis, and also
specialise to one extra dimension with the soliton being a domain wall
or kink.  The brane is thus smooth and of finite thickness.

In the standard model, scalar fields are used to spontaneously break the
electroweak gauge symmetry.  We are thus led to explore the possibility
that the scalar field composing the brane also plays a role in some internal
symmetry breaking (not necessarily electroweak).  This can be achieved
in a toy-model sense through domain wall configurations displaying
a feature dubbed the ``clash of symmetries'' \cite{clash1}, a generalization of the 
simple $Z_2$ kink.

We first briefly review the $Z_2$ kink.  Take a scalar field in a spacetime
of any dimensionality $d$ and give it the potential energy $V = \lambda(\phi^2 - u^2)^2$,
invariant under the discrete $Z_2$ symmetry $\phi \to -\phi$, and with $\lambda>0$. 
The global minima are $\phi = +u$ and $\phi = -u$.  They are degenerate and
separated by a potential barrier; the vacuum ``manifold'' is simply
two disconnected points, reflecting the spontaneous breaking of the $Z_2$
symmetry.  A kink configuration is a static, topologically-stable solution of
the Euler-Lagrange equations that depends on one of the spatial dimensions only,
and has the two global minima as asymptotic boundary conditions.
For the quartic potential case quoted, the solutions are
\begin{equation}
\phi(w) = \pm u \tanh(\sqrt{2 \lambda} u w)
\end{equation}
where $w$ is the extra-dimension coordinate and $\pm$ refers to kink and
antikink, respectively.  The kink has width $1/\sqrt{2\lambda}u$, and becomes
a step function as $\lambda \to \infty$.  It is a fuzzy
$d-1$-dimensional brane-like entity.

The clash of symmetries \cite{clash1}
generalises the above by using a vacuum manifold that
consists not just of disconnected points, but rather a set of disconnected
copies of a non-trivial manifold.  Take a Higgs model with a continuous
symmetry $G$ that spontaneously breaks to $H$ at the global minima of the potential,
and that also features a spontaneously broken discrete symmetry lying outside
of $G$.  The vacuum manifold is a set of disconnected copies of the coset
space $G/H$, with the number of copies given by the discrete symmetry
breaking pattern.  

Each point within a $G/H$ corresponds to a {\it differently embedded} $H$ subgroup.
But each such point is now a potential boundary condition for a generalized
kink configuration.  We define a kink in this context as a static 1-dimensional
solution to the scalar-field Euler-Lagrange equations that interpolates
between a given point in one $G/H$ and a certain point in a $G/H$ disconnected
from the first.  The straightforward analogue of the $Z_2$ kink requires
these two points to correspond to {\it identically embedded} subgroups $H$.
In this case, the ``instantaneous'' unbroken subgroup as a function of $w$
is always the same subgroup $H$.  A clash of symmetries kink has the two 
boundary condition points as corresponding to differently embedded (but
isomorphic) $H$ subgroups.  At finite $w$, the configuration displays an
unbroken symmetry $H_{\rm finite}$ that is smaller than $H$, typically just the intersection
$H(w = -\infty) \cap H(w = +\infty)$.  At precisely $w=0$, the odd-function
components of the Higgs configuration vanish, leading to an instantaneously
enhanced symmetry we shall call $H_0$ ({\it not} isomorphic to $H$ in general).
We have thus achieved a spatially-dependent symmetry breaking pattern.

It has been speculated that a hierarchical symmetry breaking cascade
$G \to H_0 \to H_{\rm finite}$ may be felt by degrees of freedom that are
almost $\delta$-function confined near $w=0$ \cite{shin}.  If the confinement was
precisely of $\delta$-function form, then the brane localized effective
theory would display $G \to H_0$, assuming local interactions between
the brane and bulk fields.  But given the finite thickness of the brane,
one would rather expect localization of finite width, in which case
the further breaking $H_0 \to H_{\rm finite}$ would feature in the effective
theory at a lower energy scale driven by the localization width.
No explicit realization of this idea has yet been found.  The well-known 
Yukawa-style localization of 4-d fermion zero modes,
at least in its simplest form, appears to not
be what is needed to realize this idea  \cite{curtin}.

\section{$O(10)$ kinks: breaking to $SU(3) \otimes SU(2) \otimes U(1)^2$}

I now display some explicit $O(10)$ adjoint-Higgs kinks displaying the
clash of symmetries \cite{shin}.  Writing the adjoint Higgs $\Phi$ as a $10 \times 10$
antisymmetric matrix, the most general quartic potential has the
terms $Tr(\Phi^2)$, $(Tr(\Phi^2))^2$ and $Tr(\Phi^4)$.  The invariant
cubic term $Tr(\Phi^3)$ is identically zero, so there is an accidental
$\Phi \to -\Phi$ discrete symmetry also, and it lies outside of $O(10)$.
The minimisation of this Higgs potential was studied in Ref.\cite{Li}.  The first step
is to use an $O(10)$ transformation to bring the VEV into the standard
form $\Phi = {\rm diag}(a_1\epsilon,a_2\epsilon,a_3\epsilon,a_4\epsilon,a_5\epsilon)$,
where $\epsilon = i\sigma_2$ is the $2 \times 2$ antisymmetric matrix.
For a certain range of parameters, the global minima are
\begin{equation}
a_i^2 = {\rm const}. \equiv a^2_{\rm min} \forall i,
\end{equation}
where the constant is a certain combination of Higgs potential parameters.
The unbroken subgroup is $U(5)$.

We now seek kink solutions.  Let us choose $\Phi(-\infty) = 
- a_{\rm min} {\rm diag}(\epsilon,\epsilon,\epsilon,\epsilon,\epsilon)$ as the
boundary condition at $w = -\infty$, where the
overall minus sign is related to the $\Phi \to -\Phi$ breaking.  
At $w = +\infty$ there are three sensible choices:
\begin{equation}
\Phi(+\infty) = \left\{ \begin{array}{c}
a_{\rm min}\, {\rm diag}(\epsilon,\epsilon,\epsilon,\epsilon,\epsilon) \\
a_{\rm min}\, {\rm diag}(\epsilon,\epsilon,\epsilon,-\epsilon,-\epsilon) \\
a_{\rm min}\, {\rm diag}(\epsilon,-\epsilon,-\epsilon,-\epsilon,-\epsilon)
\end{array} \right. ,
\end{equation}
giving three kinds of kinks: symmetric, asymmetric and super-asymmetric, respectively.
(There must be an odd number of relative minus signs between the $-\infty$ and
$+\infty$ boundary
conditions to ensure that they cannot be transformed into each other under $O(10)$.)
The kink configurations are of the form
$\Phi_k(w) = \alpha(w)\, \Phi(-\infty) + \beta(w)\, \Phi(+\infty)$
leading to
$H_{\rm finite} = \ U(5)$, $U(3)\otimes U(2)$, $U(4) \otimes U(1)$
respectively.  The asymmetric and super-asymmetric kinks display the clash of
symmetries phenomenon.  Since
$U(3) \otimes U(2) \cong G_{SM} \otimes U(1)'$, asymmetric $O(10)$ kinks
show some model-building promise.  For the parameter slice where the 
coefficient of the $(Tr(\Phi^2))^2$ term vanishes, there is an
analytic solution:
\begin{equation}
a_1(w) = a_2(w) = a_3(w) = a_{\rm min}\tanh(\mu w),\quad a_4(w) = a_5(w) = -a_{\rm min},
\end{equation}
where $1/\mu$ is the width of the domain wall.  For this parameter slice, the
super-asymmetric kink has the lowest energy density.  The symmetric and
asymmetric kinks would therefore be expected to be unstable to evolution to
the lowest energy configuration.  It is unknown whether or not the asymmetric
configuration has the lowest energy density in other regions of parameter
space, where the functional forms of $a_{1,2,3}(w)$ and $a_{4,5}(w)$ would
be different from the simple analytic solution quoted above.  There is also
no a priori need to stick with a quartic potential.

Notice that at $w=0$, the odd-functions $a_{1,2,3}$ vanish, leading to
$H_0 = O(6) \otimes U(2) \cong SU(4) \otimes SU(2) \otimes U(1)$.  It would
be nice to find a way to realise the hierarchical breaking cascade
\begin{equation}
O(10) \to SU(4) \otimes SU(2) \otimes U(1) \to G_{SM} \otimes U(1)'
\end{equation}
as per the speculative idea discussed previously.

\section{Randall-Sundrum gravity}

Consider a $U(1) \otimes U(1)$ model with two complex Higgs fields $\Phi_{1,2}$
minimally coupled to 5-d Einstein gravity.  Adopting the Randall-Sundrum style
warped metric ansatz $ds_5^2 = dw^2 + e^{2f(w)} ds_4^2$ with a Minkowski brane \cite{rs1,rs2},
the coupled Einstein-Higgs equations yield the solution
\begin{equation}
\Phi_{1,2}(w) = \frac{u}{\sqrt{2}}\sqrt{1 \pm \tanh\beta w},\ \ 
f(w) = - \frac{u^2}{12\kappa} \ln (\cosh\beta w)
\end{equation}
for a certain sextic Higgs potential \cite{clash2}.  The parameter $u$ is a VEV, 
$\beta$ is the inverse domain wall width, while $\kappa$ is
related to the 5-d Planck mass.  The warp factor exponent is a smooth analogue of the 
$-k|w|$ familiar from the RS2 $\delta$-function brane case.  An RS2-like \cite{rs2}
limit is reached when $\beta \to \infty$ and $u \to 0$ such that $u^2\beta$ is kept
finite.  Notice that this is a strange limit in this context, because the kink
amplitude $u$ shrinks to zero.

The $\Phi_{1,2}$ kinks display a primordial clash of symmetries feature, because
asymptotically a different $U(1)$ group is exact on opposite sides of the wall,
while both $U(1)$'s are broken at all finite $w$.  This construction shows that
the clash of symmetries is compatible with a smoothed out version of RS2 gravity.

\section{Prospects, issues, questions}

Ultimately, one would want to construct a Higgs kink brane-world model with
a phenomenologically successful 4-d effective standard model or extension
thereof on the domain wall.  Hopefully the clash of symmetries or a related
idea could be used to connect brane formation with at least some of
the required internal symmetry breaking, for example $SO(10)$ GUT breaking.
The resulting model should also have a successful 4-d effective cosmology (see
\cite{slatyer} and references therein for an introduction to kink-brane cosmology).

To achieve this end, we need to simultaneously localize fermions, gravitons \cite{rs2}, 
gauge bosons \cite{ds} and possibly also some Higgs bosons to the domain wall.  Mechanisms
exist for all these disparate fields, but they need to be non-trivially combined
so as to yield successful particle and cosmological 4-d phenomenology.
This is an interesting prospect and challenge.


\begin{theacknowledgments}
I thank the organisers for the invitation, and my collaborators
on brane-world research: D. Curtin, G. Dando, A. Davidson, A. Demaria, D. George, J. Rozowsky,
E. Shin, T. Slatyer, B. Toner and K. Wali.  This work was supported by the Australian 
Research Council.
\end{theacknowledgments}



\bibliographystyle{aipproc}   




\end{document}